\documentclass{aa}  
\usepackage{graphicx}
\usepackage{natbib}
\bibpunct{(}{)}{;}{a}{}{,}
\usepackage{txfonts}
\begin{document}
   \title{No evidence of a hot Jupiter around HD\,188753\,A\thanks{Based on observations collected at the Observatoire de Haute-Provence 
	  with the ELODIE echelle spectrograph mounted on the 1.93-m telescope.}}

   \author{A.~Eggenberger\inst{1} \and S.~Udry\inst{1} \and T.~Mazeh\inst{2}
   \and Y.~Segal\inst{2} \and M.~Mayor\inst{1}}

   \offprints{Anne Eggenberger, \email{Anne.Eggenberger@obs.unige.ch}}

   \institute{Observatoire de Gen\`eve, 51 ch. des Maillettes, 
             1290 Sauverny, Switzerland
         \and
	     School of Physics and Astronomy, Raymond and Beverly Sackler 
	     Faculty of Exact Sciences, Tel Aviv University, Tel Aviv, Israel  
             }

   \date{Received / Accepted }

 
  \abstract
   {The discovery of a short-period giant planet (a hot Jupiter) around the 
   primary component 
   of the triple star system HD\,188753 has often been considered as an 
   important observational evidence and as a serious challenge to 
   planet-formation theories.}
   {Following this discovery, we monitored HD\,188753 during one
   year to better characterize the planetary orbit and the 
   feasibility of planet searches in close binaries and multiple star systems.}
   {We obtained Doppler measurements of HD\,188753 with the ELODIE spectrograph 
   at the Observatoire de Haute-Provence. We then extracted radial velocities 
   for the two brightest components of the system using our multi-order, two-dimensional
   correlation algorithm, TODCOR.}
   {Our observations and analysis do not confirm the existence of the 
   short-period giant planet previously reported around HD\,188753\,A.  
   Monte Carlo simulations show that we had both the precision and the temporal
   sampling required to detect a planetary signal like the one quoted.}
   {From our failure to detect the presumed planet around HD\,188753\,A and
   from the available data on HD\,188753, we conclude that there 
   is currently no convincing evidence of a close-in giant planet around 
   HD\,188753\,A.}

   

   \maketitle

%

\section{Introduction}
\label{introduction}

In a recent paper, \citet{Konacki05} has reported the discovery of a
3.35-day radial-velocity modulation of \object{HD\,188753\,A} with a semiamplitude
of 149~m\,s$^{-1}$, and attributed this signal to the presence of a planet 
with a minimum mass of 1.14\,M$_{\rm Jup}$ in orbit around the star. 
Interestingly, HD\,188753\,A is the primary component of a  
triple star system, while the visual companion, \object{HD\,188753\,B}, is itself 
a spectroscopic binary with a period of 155~days 
\citep{Griffin77,Konacki05}. The visual orbit of the AB pair 
is characterized by a period of 25.7~years, a semimajor axis of 12.3~AU
(0.27\arcsec\ separation), and an eccentricity of 0.5 \citep{Soederhjelm99}. 
\object{HD\,188753} is therefore a hierarchical system, with the primary component 
(A) hosting the giant planet and the secondary (Ba) and tertiary (Bb) 
components forming a close pair in orbit at a distance of 
12.3~AU from the planet-host star. 

The discovery of a giant planet around HD\,188753\,A has attracted much
attention, as the proximity of HD\,188753\,B poses a serious problem 
for planet formation theories. Indeed, the periastron distance 
of the visual binary is only 6.2~AU, so that any protoplanetary disk
around HD\,188753\,A would be truncated at $\sim$2~AU 
\citep{Konacki05,Jang-Condell07}, i.e. probably inside the snow line. 
The favored core accretion model, which stipulates that the
cores of giant planets form beyond the snow line, might thus have a
problem accounting for the existence of a giant planet around 
HD\,188753\,A. Assuming that the system has maintained its current
orbital configuration ever since the planet formed, explaining the existence
of a hot Jupiter around the primary component of HD\,188753 remains a challenge
\citep{Nelson00,Mayer05,Boss06,Jang-Condell07} and might indicate
that the core accretion mechanism is not able to account for the
existence of all the planets discovered so far.

An alternative way to explain the existence of a close-in Jovian planet 
around HD\,188753\,A and also to circumvent the theoretical
problem is to assume that the current orbital configuration is the result of a
dynamical process \citep{Pfahl05,PortegiesZwart05}. According to this approach, 
HD\,188753\,A was a single star or had a more 
distant stellar companion at the planet formation phase, and this initial state
was later transformed through a dynamical encounter into the triple
system we presently observe. 
\citet{Pfahl06} estimate that dynamical interactions could deposit 
giant planets in about 0.1\% of the binaries with a semimajor axis  
$a$\,$<$\,50~AU. The existence of a short-period Jovian planet around 
HD\,188753\,A might thus be the result of a dynamical process, rather than 
of an in situ formation 
\citep{Pfahl05,PortegiesZwart05,Jang-Condell07}.

Among the $\sim$$200$ extrasolar planets discovered so far by Doppler spectroscopy, 33 
are known to orbit the components of binaries or triple stars
\citep{Eggenberger04,Mugrauer05b,Raghavan06}. However, unlike the
giant planet around HD\,188753\,A, most of these planets reside in
systems with separations larger than 100~AU. Since double stars closer
than 2--7\arcsec\ present observational difficulties for Doppler
studies, they have commonly been left out of radial-velocity planet
searches. The occurrence of planets in binaries with semimajor axes below
100~AU is therefore still largely unprobed. Despite this strong bias and 
prior to the discovery by Konacki, three planets were found in
stellar systems with separations close to 20~AU: Gliese\,86
\citep{Queloz00,Els01,Mugrauer05,Lagrange06}, $\gamma$\,Cephei
\citep{Hatzes03,Neuhaeuser07,Torres07},
and HD\,41004 \citep{Zucker03,Zucker04}. Two features render the planet around
HD\,188753\,A particularly interesting. First, both the
planetary and the binary orbital parameters are known (only
$\gamma$\,Cephei shares this property). Second, the binary
periastron distance may be small enough to preclude giant planet
formation according to the canonical models.

Doppler searches for planets in close binaries have recently proven
feasible using dedicated reduction techniques based on two-dimensional
correlation \citep{Zucker03,Konacki05b}. In order to probe the occurrence of 
planets in close stellar systems, two surveys searching for planets
in spectroscopic binaries are currently underway
\citep{Eggenberger02,Konacki05b}, the planet around HD\,188753\,A
being a product of Konacki's survey. Yet, deriving the 
velocity of HD\,188753\,A to the precision needed to reveal the
presence of a planet is particularly challenging, as the
radial velocity of the secondary, HD\,188753\,Ba, varies with a 
semiamplitude of $\sim$$13$~km\,s$^{-1}$ over a timescale of 155~days.

In order to study this intriguing system further, we monitored  
HD\,188753 during one year with the ELODIE spectrograph 
\citep{Baranne96}. We then used our multi-order 
TODCOR algorithm \citep{Zucker03} to derive the radial velocities of
the two brightest components, namely HD\,188753\,A and HD\,188753\,Ba. Unfortunately, our data 
and analysis do not confirm the existence of the planet reported by 
\citet{Konacki05} 
around HD\,188753\,A. We present our observations and data reduction 
technique in Sect.~\ref{observations}. Our results and the lack 
of evidence of a hot Jupiter around HD\,188753\,A 
are described in Sect.~\ref{results} and discussed in Sect.~\ref{discussion}.

\section{Observations and data analysis}        
\label{observations}                            

We observed HD\,188753 with the ELODIE echelle spectrograph
\citep{Baranne96} at the Observatoire de Haute-Provence (France)
between July 2005 and August 2006. Altogether, we gathered 48 spectra
with a typical signal-to-noise ratio of 55 (per pixel at 550~nm). Note
that ELODIE is a fiber-fed spectrograph with a fiber of 2\arcsec\ in 
diameter. As the angular separation between HD\,188753\,A and
HD\,188753\,B is always less than 0.4\arcsec, our observations
necessarily record the combined spectrum of the whole system. In the
course of the observations, we realized that the precision seemed
degraded when the two brightest stars (A and Ba) had similar velocities 
(velocity difference of less than a few km\,s$^{-1}$), and we subsequently
avoided observing the system at those particular phases of the 
155-day modulation.

Our spectra were reduced and cross-correlated online, the wavelength 
calibration being
provided by the thorium-argon simultaneous reference technique
\citep{Baranne96}. The double-lined nature of HD\,188753 was evident
at the telescope, where two blended correlation features could 
be seen, one corresponding to component A, the other corresponding to 
component Ba. 
The contribution of the third component (Bb) to the total flux is quite 
modest, so that the system can basically be considered as a double-lined 
spectroscopic binary. Nonetheless, extracting precise radial velocities for 
the two brightest stars of the system is challenging, for two main reasons: 
(i) the difference in mean velocity between components A and Ba is
currently only of 2~km\,s$^{-1}$; (ii) the radial
velocity of component Ba varies with a period of 155~days and a
semiamplitude of $\sim$$13$~km\,s$^{-1}$, which is not very different from the sum 
of the intrinsic widths of the two correlation features. This leads to a
situation where the two correlation features are always strongly
blended, the blend changing continuously due to the 155-day modulation.

Given the double-lined nature of HD\,188753 and the strong line
blending, we derived radial velocities using a multi-order version
\citep{Zucker03} of the two-dimensional
correlation algorithm TODCOR \citep{Zucker94}. This algorithm uses two
templates with a given flux ratio and unknown Doppler shifts to
compute the two-dimensional correlation function, whose maximum simultaneously 
gives the radial velocity of both components. Our version of 
TODCOR uses as templates high signal-to-noise stellar spectra built up
using spectra from our planet search programs with CORALIE and ELODIE
\citep{Queloz00,Perrier03}, along with spectra from the surveys for
low-mass companions to M dwarfs by Delfosse and coworkers
\citep{Delfosse98,Delfosse99}. Each template can furthermore be
convolved with a rotational broadening profile, thus allowing for a
better fit to the observed spectrum. In the multi-order version of TODCOR, the 
flux ratio is a function of
wavelength and is calculated for each order according to the spectral
types of the two templates using the library of spectral energy
distributions by \citet{Pickles98}. To insure fine-tuning with the
observed system, the table of flux ratios can also be
multiplied by a global normalization factor.

\begin{figure*}
\centering
\resizebox{\hsize}{!}{\includegraphics{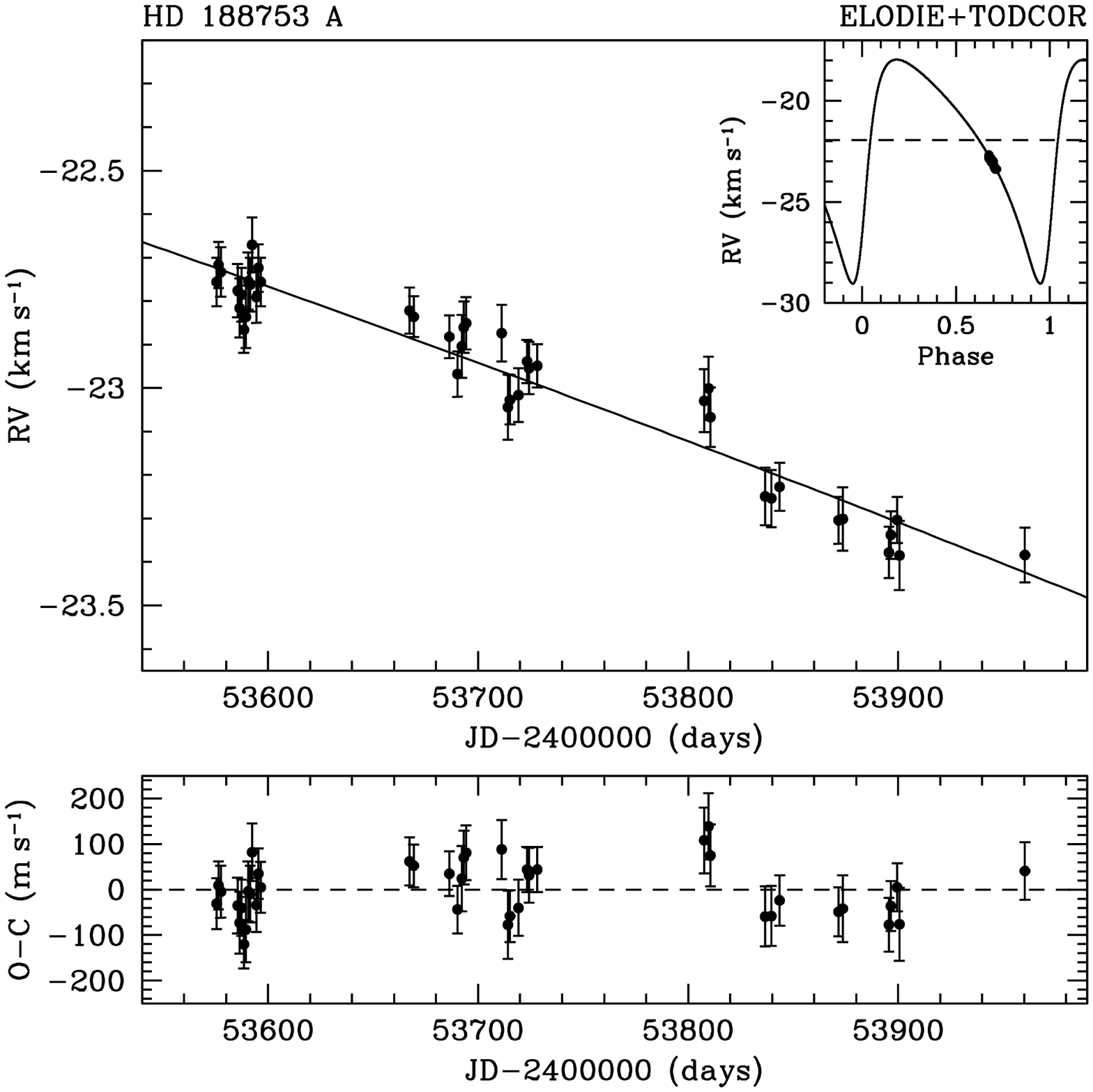}
\includegraphics{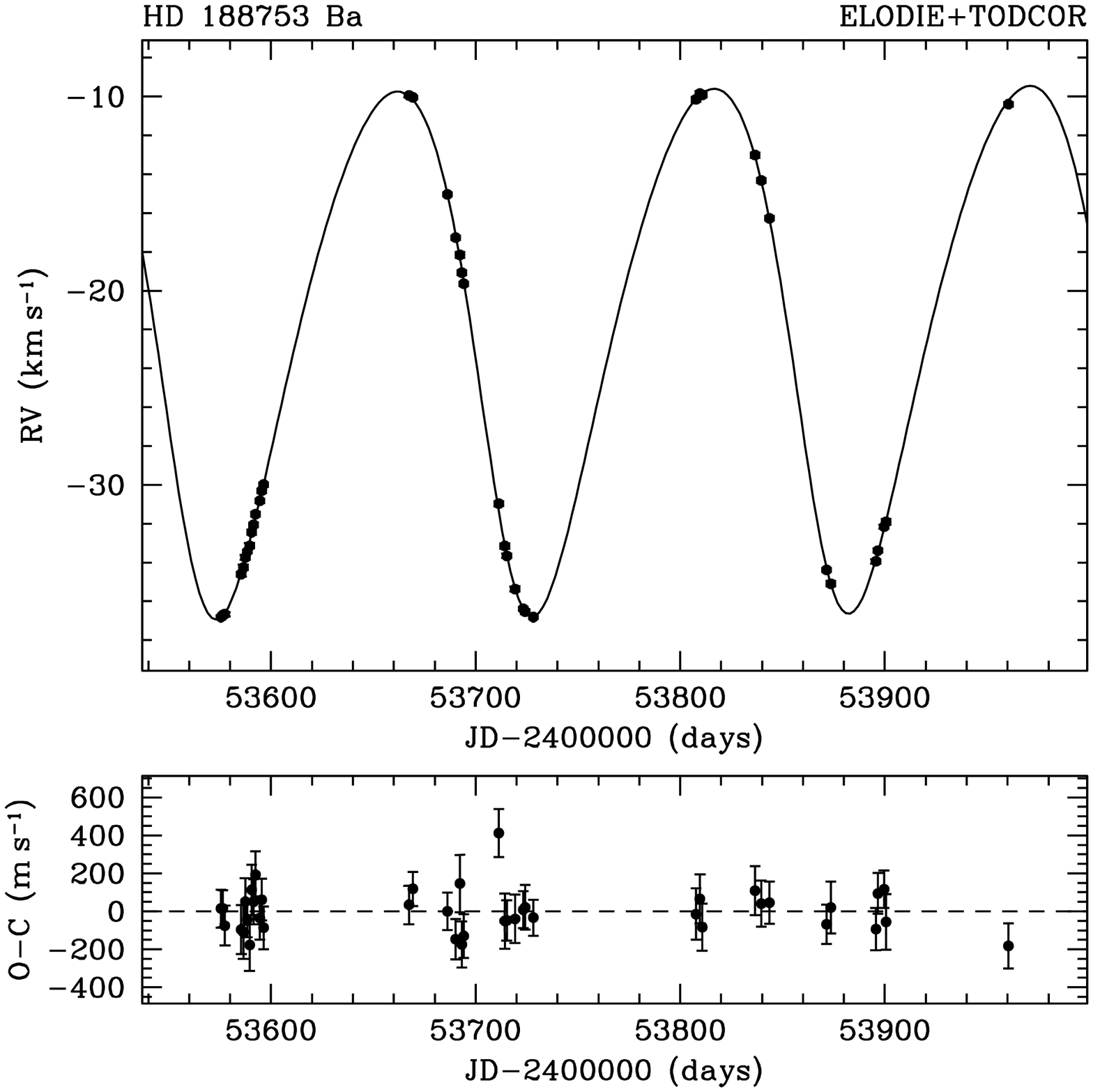}}
\caption{Radial velocities and orbital solutions for HD\,188753\,A
(left) and HD\,188753\,Ba (right). For component A, the solid line
represents the 25.7-year orbital motion of the visual pair shown in
full in the inset. This orbit was computed by using 
the elements in common with the visual orbit as fixed parameters 
\citep{Soederhjelm99},
using our radial velocities (shown as dots) to fit only the systemic
velocity and the radial-velocity semiamplitude. The residuals around
this long-period orbit are shown in the bottom panel. The rms is
60~m\,s$^{-1}$. The best-fit orbital solution for component Ba is
given in Table~\ref{tab2}, and it includes a linear drift to take 
the long-period orbital motion of the AB pair into account. The bottom panel 
shows the 
residuals around this solution (rms of $107$~m\,s$^{-1}$).}
\label{fig1}
\end{figure*}

To derive the radial velocities of HD\,188753\,A and Ba, we ran TODCOR
for a variety of different pairs of templates from our library,
searching for the pair that best matches our observed composite
spectra. The two templates finally retained were the spectrum of 
HD\,224752 (a G6 dwarf) for HD\,188753\,A and the spectrum of 
HD\,225208 (a K0 dwarf) for HD\,188753\,Ba. 
This template configuration was chosen because (i) it gave the
lowest residuals for the spectroscopic orbit of component Ba, and  
(ii) the primary template maximized the correlation
coefficient for component A. Additional fine-tuning consisted of
broadening the primary and secondary templates by $\upsilon\sin{i}$ of
$1$~km\,s$^{-1}$ and $3$~km\,s$^{-1}$, respectively, on
top of their initial broadening of $3.6$~km\,s$^{-1}$ and
$1.4$~km\,s$^{-1}$.  With this TODCOR setup, a third correlation
feature corresponding to component Bb was visible at some orbital
phases. Although not feasible with TODCOR, extracting the exact radial
velocity of component Bb from our spectra is possible and will be the subject
of a forthcoming paper (Mazeh et al. in prep.).

\section{Results}
\label{results}

Our radial velocities for HD\,188753\,A and HD\,188753\,Ba are
displayed in Fig.~\ref{fig1} and are listed in Table~\ref{tab1} 
available in the electronic version. 
Internal uncertainties for each measurement were first
estimated using a one-dimensional error analysis as explained in
\citet{Zucker94}. Internal uncertainties computed in this way seem
correct from a relative point of view, but are too pessimistic in
absolute terms when compared to external uncertainties measured by the
root-mean-squares (rms) around the orbits. This is true not only for
HD\,188753\,A and Ba, but also for the few other systems we have
analyzed so far with TODCOR. We consequently used the one-dimensional
uncertainties as relative weights, but then rescaled them globally,
forcing the match with the rms derived from the external error
analysis. The internal uncertainties finally used (and given in
electronic form) are those rescaled values.  

\onltab{1}{
\begin{table*}
\caption{Radial velocities for HD\,188753\,A and HD\,188753\,Ba.}
\centering
\begin{tabular}{ccccc}
\hline\hline
Julian date       & Radial velocity            & Uncertainty     & Radial velocity      & Uncertainty    \\
                  & of HD\,188753\,A           &                 & of HD\,188753\,Ba    &      \\
(JD-2\,400\,000)  & (km\,s$^{-1}$)             & (km\,s$^{-1}$)  &(km\,s$^{-1}$)        & (km\,s$^{-1}$) \\
\hline
53575.47219    &    $-$22.756    &    0.056     &    $-$36.815    &   0.100   \\
53576.46953    &    $-$22.717    &    0.053     &    $-$36.711    &   0.098   \\
53577.47838    &    $-$22.733    &    0.057     &    $-$36.662    &   0.105   \\
53585.44767    &    $-$22.776    &    0.061     &    $-$34.600    &   0.129   \\
53586.45691    &    $-$22.816    &    0.068     &    $-$34.251    &   0.141   \\
53587.40502    &    $-$22.785    &    0.062     &    $-$33.734    &   0.123   \\
53588.43320    &    $-$22.866    &    0.053     &    $-$33.427    &   0.093   \\
53589.48392    &    $-$22.836    &    0.072     &    $-$33.138    &   0.135   \\
53590.46733    &    $-$22.754    &    0.066     &    $-$32.440    &   0.133   \\
53591.50460    &    $-$22.762    &    0.062     &    $-$32.056    &   0.119   \\
53592.42986    &    $-$22.671    &    0.063     &    $-$31.515    &   0.123   \\
53594.48743    &    $-$22.791    &    0.059     &    $-$30.826    &   0.114   \\
53595.41994    &    $-$22.724    &    0.055     &    $-$30.305    &   0.110   \\ 
53596.44843    &    $-$22.756    &    0.056     &    $-$29.980    &   0.112   \\
53667.37993    &    $-$22.822    &    0.053     &    $-$9.933     &   0.101   \\
53669.29918    &    $-$22.836    &    0.047     &    $-$10.044    &   0.089   \\
53686.25752    &    $-$22.882    &    0.049     &    $-$15.046    &   0.099   \\
53690.26372    &    $-$22.968    &    0.052     &    $-$17.270    &   0.106   \\
53692.26624    &    $-$22.904    &    0.072     &    $-$18.145    &   0.151   \\
53693.25587    &    $-$22.860    &    0.059     &    $-$19.072    &   0.122   \\
53694.23432    &    $-$22.851    &    0.060     &    $-$19.647    &   0.116   \\
53711.29621    &    $-$22.874    &    0.065     &    $-$30.974    &   0.126   \\
53714.24526    &    $-$23.044    &    0.075     &    $-$33.152    &   0.145   \\
53715.21499    &    $-$23.027    &    0.057     &    $-$33.653    &   0.106   \\
53719.21177    &    $-$23.016    &    0.062     &    $-$35.359    &   0.128   \\
53723.24701    &    $-$22.939    &    0.050     &    $-$36.383    &   0.098   \\
53724.23431    &    $-$22.954    &    0.060     &    $-$36.528    &   0.117   \\
53728.23107    &    $-$22.949    &    0.050     &    $-$36.814    &   0.095   \\
53807.68889    &    $-$23.029    &    0.072     &    $-$10.138    &   0.135   \\
53809.68173    &    $-$23.001    &    0.073     &    $-$9.852     &   0.129   \\
53810.67510    &    $-$23.067    &    0.068     &    $-$9.917     &   0.124   \\
53836.63984    &    $-$23.249    &    0.066     &    $-$13.014    &   0.130   \\
53839.62754    &    $-$23.254    &    0.066     &    $-$14.314    &   0.121   \\
53843.60439    &    $-$23.227    &    0.055     &    $-$16.263    &   0.111   \\
53871.58461    &    $-$23.304    &    0.054     &    $-$34.381    &   0.104   \\
53873.59403    &    $-$23.301    &    0.073     &    $-$35.092    &   0.136   \\
53895.59737    &    $-$23.378    &    0.059     &    $-$33.937    &   0.111   \\
53896.58051    &    $-$23.338    &    0.055     &    $-$33.379    &   0.108   \\
53899.58774    &    $-$23.303    &    0.053     &    $-$32.146    &   0.098   \\
53900.55889    &    $-$23.385    &    0.080     &    $-$31.905    &   0.146   \\
53960.38404    &    $-$23.383    &    0.063     &    $-$10.397    &   0.118   \\
\hline
\end{tabular}
\label{tab1}
\end{table*}
}

The error analysis shows that the precision on our radial velocities
is degraded (factor of about 2) when the velocity difference between 
components A and Ba is less
than 3~km\,s$^{-1}$. Note that, although not reported, Konacki probably
faced the same problem, since in his measurements, taken in 7 runs spread over
466~days, the difference in
radial velocity between the two brightest components is always greater 
than 7~km\,s$^{-1}$, which is not likely to be a coincidence as this happens 
only
about half of the time. In consequence, we derived orbital solutions using
only our 41 measurements with a velocity difference larger than
3~km\,s$^{-1}$.

\begin{figure*}
\centering
\resizebox{\hsize}{!}{\includegraphics{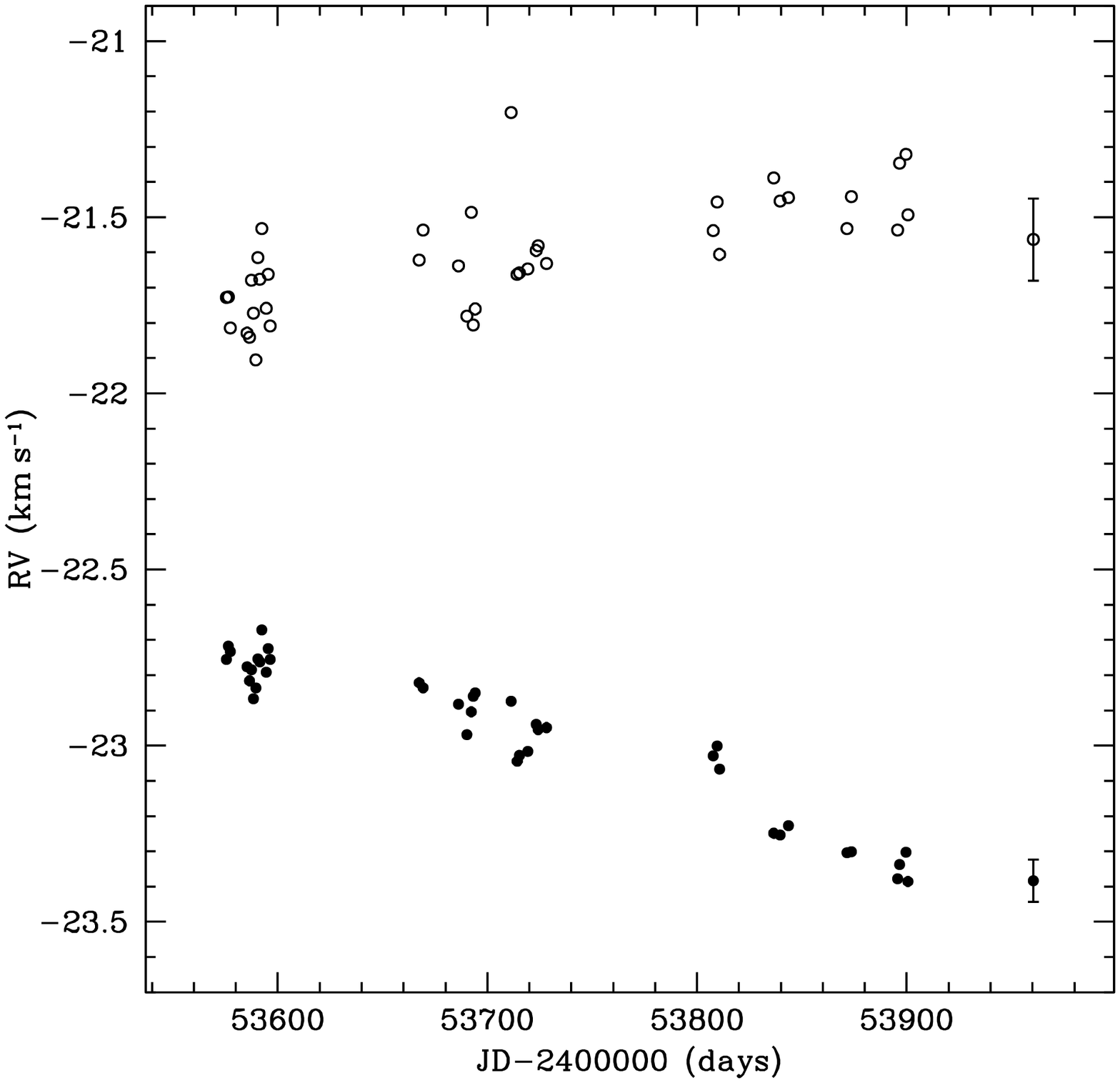}
\includegraphics{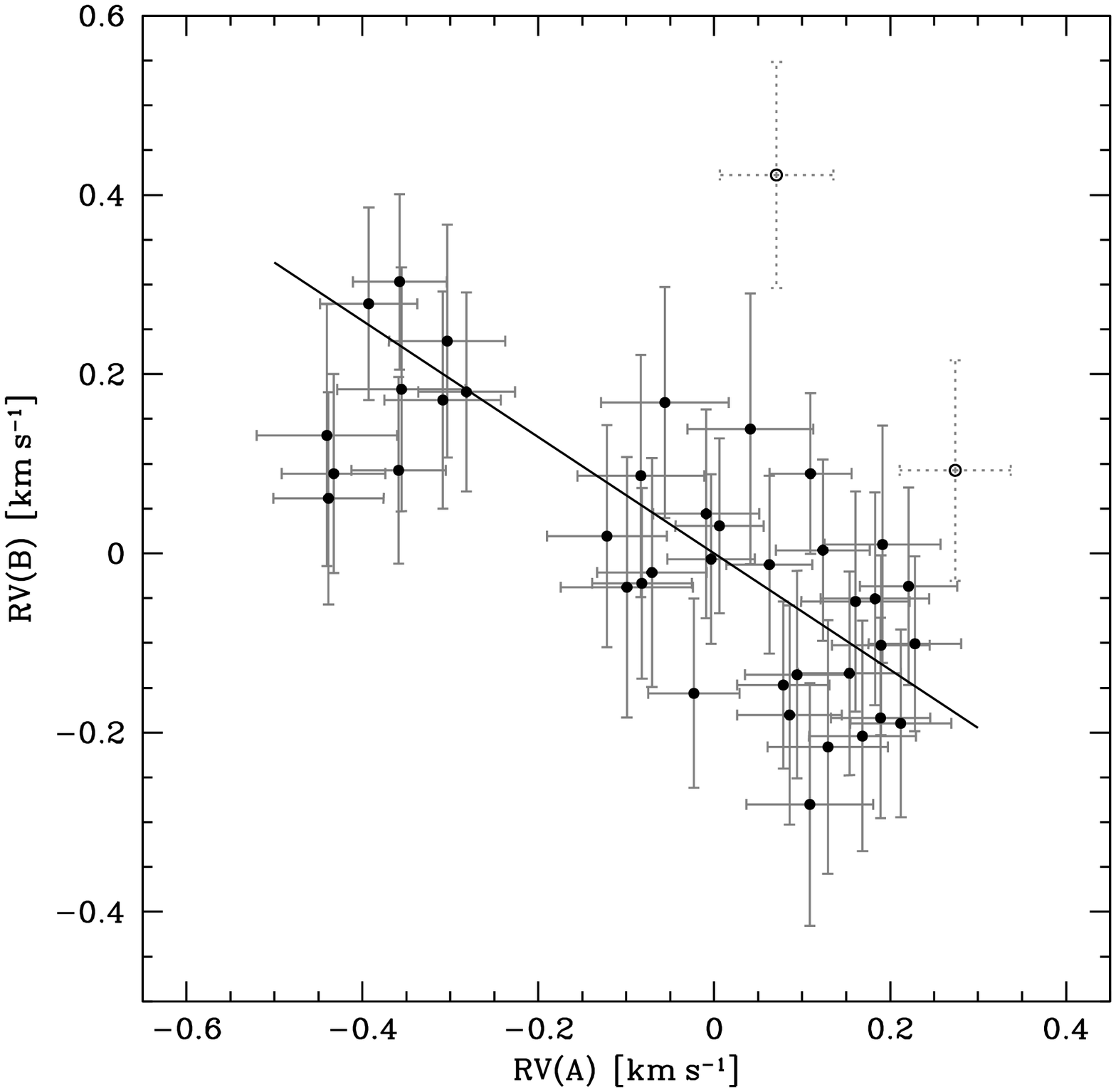}}
\caption{{\it Left:} Radial velocities for HD\,188753\,A (dots) and 
for HD\,188753\,Ba after having removed the 155-day modulation (open circles). 
For the sake of clarity, only a typical error bar is displayed on the last
measurement of each component. {\it Right:} Wilson-Mazeh diagram showing the 
radial velocities of HD\,188753\,Ba (as displayed on the 
left panel but with the average velocity subtracted) versus the radial velocities 
of HD\,188753\,A (again with the average velocity subtracted). As expected, our 
measurements are consistent with a straight line, except for two somewhat 
discordant points represented by open symbols and dotted error bars. 
The straight line shown on this plot is the result of a least-square 
fit taking individual uncertainties on both axes into account and ignoring the two
discordant measurements. The slope of this best-fit line yields a mass ratio 
$q_{AB}$\,$=$\,$1.54\pm0.19$.}
\label{wilson-mazeh}
\end{figure*}

As shown by \citet{Konacki05}, the spectroscopic binary with a 
period of 155~days detected by \citet{Griffin77} is the B component. 
Our single-lined spectroscopic orbit for
HD\,188753\,B (including a linear drift to take the
long-period AB motion into account) is displayed in Fig.~\ref{fig1}. The corresponding
orbital elements are listed in Table~\ref{tab2}. All these elements
are relatively well constrained. 

For the A component, the dominant motion seen in our data is a steady 
decrease in velocity. To check the consistency of this velocity decrease 
with the 25.7-year orbital motion of the AB pair, we fitted our radial 
velocities with a Keplerian curve computed using the elements from the 
visual orbit \citep{Soederhjelm99} as fixed parameters. The only two free
parameters in the fit were thus the systemic velocity and the velocity 
semiamplitude, $K_A$. The result is displayed in Fig.~\ref{fig1} and shows
that the observed velocity decrease is indeed consistent with the
orbital motion of the AB pair. The fit in Fig.~\ref{fig1} 
yields a velocity semiamplitude 
$K_A$\,$=$\,$5.54$\,$\pm$\,$0.25$~km\,s$^{-1}$. Combining
this value with the sum 
$K_A$\,$+$\,$K_B$\,$=$\,$9.23$\,$\pm$\,$0.86$~km\,s$^{-1}$ derived
from the elements of the visual orbit\footnote{\citet{Soederhjelm99} does
not give uncertainties on the orbital elements. Our calculation assumes
uncertainties of 0.1~years on the period, of 0.02\arcsec\ on the angular
semimajor axis, and of 0.03 on the eccentricity.}, we obtain a mass ratio 
$q_{AB}$\,$=$\,$M_B/M_A$\,$=$\,$1.50\pm0.37$. This value is fully consistent 
with the masses reported by \citet{Konacki05}, which yield a
mass ratio $q_{AB}$\,$=$\,$1.54\pm0.12$. 

\begin{table}
\caption{Orbital parameters for the 155-day spectroscopic orbit of 
HD\,188753\,Ba, where the long-period orbital motion of the AB pair was taken 
into account by including a linear drift.}
\begin{center}
\begin{tabular}{llc}
\hline\hline
Parameter & Units & Value \\
\hline
$P$                         & (days)            & $154.56\pm0.08$\\
$T$                         & (JD-2400000)      & $53405.98\pm0.51$\\
$e$                         &                   & $0.1671\pm0.0017$\\
$\gamma$                    & (km\,s$^{-1}$)     & $-21.625\pm0.034$\\
$\omega$                    & (deg)             & $136.2\pm1.2$\\
$K$                         & (km\,s$^{-1}$)     & $13.554\pm0.029$\\
Linear drift                & (km\,s$^{-1}$\,yr$^{-1}$)  & $0.342\pm0.071$ \\
$a_1\sin{i}$                & (AU)              & $0.18986\pm0.00042$ \\
$f(m)$                      & (M$_{\sun}$)      & $0.03822\pm0.00025$  \\
\cline{1-3}
$N_{\rm meas}$              &                   & 41\\
rms                         & (m\,s$^{-1}$)     & 107\\
\hline
\end{tabular}
\end{center}
\label{tab2}
\end{table}

In order to further study the radial-velocity variation induced by the
wide orbit, we show in Fig.~\ref{wilson-mazeh} the velocity of 
HD\,188753\,A along with the velocity of HD\,188753\,Ba after having
removed the 155-day modulation. This is presented in two panels. The 
left-hand panel depicts the two sets of velocities as a function of time, 
where the opposite slopes of the two stars can be seen easily. The 
right-hand panel shows a Wilson-Mazeh plot \citep{Wilson41,Mazeh02} 
in which the velocities of HD\,188753\,Ba and HD\,188753\,A are plotted
against each other (with the average velocity removed from each data
set). As explained in \citet{Wilson41}, plotting the pairs of velocities
corresponding to each observation in this way should result in a straight
line, whose negative slope is the inverse of the mass ratio. 
Figure~\ref{wilson-mazeh} shows that our data points are indeed consistent with a 
straight line, except for two somewhat discordant points. Fitting the plot in 
Fig.~\ref{wilson-mazeh} with a linear relation through the origin, 
taking individual uncertainties on both axes into account and ignoring the two
discordant measurements, we obtain a mass ratio $q_{AB}$\,$=$\,$1.54\pm0.19$. 
This result is in
very good agreement with the two values reported previously, emphasizing the consistency
between our radial velocities, the visual orbit determined by 
\citet{Soederhjelm99}, and the masses derived by \citet{Konacki05}.
 
\begin{figure}
\centering
\resizebox{\hsize}{!}{\includegraphics{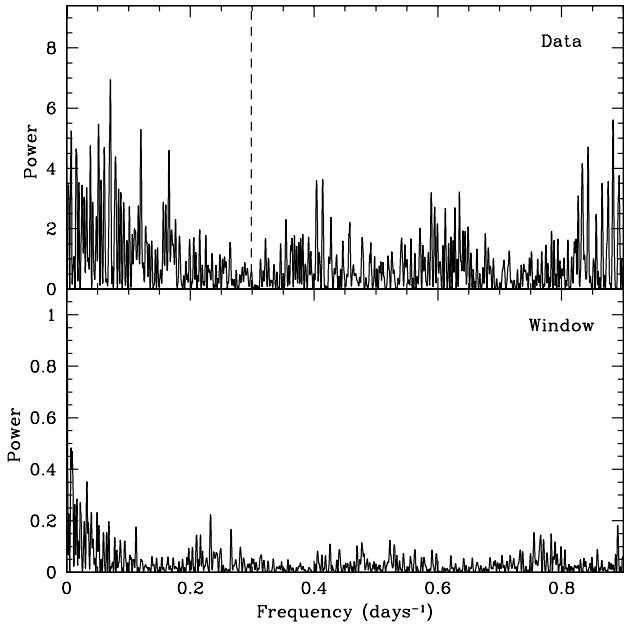}}
\caption{Lomb-Scargle periodogram of the residuals around the
  25.7-year orbital motion for HD\,188753\,A. The bottom panel shows
  the corresponding window function. The power is a measure of the
  statistical significance of a periodic signal at a given
  frequency. For this data set, the 1\% false alarm probability
  corresponds to a power of 9.4 and is represented by the top of the
  box.  The highest peak at $f$\,$=$\,0.0707~days$^{-1}$ has a false
  alarm probability of 28\%. The dashed line denotes the
  frequency of the planetary signal reported by \citet{Konacki05}.}
\label{lomb}
\end{figure}

To search for evidence of the planet found by \citet{Konacki05}, we
computed the Lomb-Scargle periodogram of the residuals around the
long-period AB orbit for component A. This is shown in
Fig.~\ref{lomb}. Contrary to our expectations, we did not find any 
significant signal in this periodogram, in particular at a frequency
of 0.30~days$^{-1}$, corresponding to the 3.35-day planet 
reported by \citet{Konacki05}. That is, our velocities for HD\,188753\,A 
show no sign of a short-period signal in addition to the 
velocity decrease related to the AB orbital motion. The rms of 60 
m\,s$^{-1}$ is basically noise and can be interpreted as the precision we achieve 
on the radial velocity of this star. 

Could it be that we missed the planet 
discovered by \citet{Konacki05}, either because of inadequate temporal 
sampling or because of insufficient precision? 
To check the adequacy of our precision and temporal sampling to detect the 
potential hot Jupiter around HD\,188753\,A, we constructed 1000 artificial
velocity sets by adding a residual value (drawn at random from our
residual data set for component A) to a planetary signal computed from
the orbital parameters quoted by \citet{Konacki05}, and sampled at our
own observing epochs. Note that in the planetary signal we included the  
linear and quadratic trends reported by \citet{Konacki05} and corresponding to 
the 25.7-year orbital motion. We then analyzed each of these artificial
velocity sets by subtracting a linear drift representing the long-period AB
orbital motion and by computing the Lomb-Scargle periodogram of the
residuals. All the mock residual sets displayed an rms
$\geq$105~m\,s$^{-1}$, that is, about twice larger than the value
actually observed. In the Lomb-Scargle periodogram, all the
artificial sets displayed a marked peak at the expected frequency,
with a false alarm probability $\leq$\,2\%. As we found the inserted
periodicity in all our 1000 simulations, there is less
than a 0.1\% probability than the planet discovered by \citet{Konacki05}
could hide in our data set. These results confirm that we indeed have 
an adequate temporal sampling and sufficient precision for detecting a
planetary signal similar to the one reported by
\citet{Konacki05}. From our failure to detect such a signal, we
conclude that our data show no evidence of a short-period massive 
planet orbiting HD\,188753\,A.

\section{Discussion and conclusions}
\label{discussion}

Our observations of HD\,188753 confirm that this system is triple and
that HD\,188753\,B is itself a spectroscopic binary with a period of
155~days. Our orbital parameters for HD\,188753\,Ba are similar to
those reported by \citet{Konacki05}, and both analyses indicate that
the B component is more massive than component A. However, we failed
to detect the presumed short-period planetary companion to
HD\,188753\,A, despite our clear ability to do so. Instead, our data
and analysis show only a steady drift in radial velocity,
consistent with the 25.7-year orbital motion of the visual pair.

Although we disagree with \citet{Konacki05} regarding the existence of
a hot Jupiter orbiting HD\,188753\,A, Konacki's velocities are
consistent with our analysis and conclusions, provided we admit that the
precision on these velocities is not 15~m\,s$^{-1}$, but rather about
140~m\,s$^{-1}$. Moreover, it should be noted that Konacki's
velocities do not support the planetary hypothesis any more than do our
own data. Indeed, when plotting a Lomb-Scargle periodogram of the residuals
of his measurements around the quadratic drift he quotes, the peak 
at 3.35~days has a false alarm probability of 24\%, and so is not 
significant. Even if the planetary solution reported by
\citet{Konacki05} satisfactorily fits his 11 velocity points (9 epochs), we
consider that this model is very likely an artificial high-frequency
fit to a small data set with large residuals. We thus conclude that there 
is currently no convincing evidence of a close-in giant planet around 
HD\,188753\,A.

The example of HD\,188753 shows that extracting precise radial velocities for 
the primary component of a spectroscopic binary can be challenging and that 
estimating the precision we really achieve on these velocities is even harder. 
Using our multi-order TODCOR algorithm that is specially designed to detect 
faint secondaries,  
we typically achieve a precision of 10--30~m\,s$^{-1}$ on the radial velocity 
of the primary component of a double or triple system with a magnitude
difference {$3$\,$<$\,$\Delta V$\,$<$\,$5.5$}; 
see \citet{Zucker04} or \citet{Eggenberger06} for examples. 
In this respect, the precision of 60~m\,s$^{-1}$ obtained here for 
HD\,188753\,A seems abnormally poor. 
One possible reason may be the lower magnitude difference
($\Delta V$\,$=$\,$0.75$) coupled to the 
large-amplitude 155-day modulation induced by component Ba. Another possibility 
may be the presence of the spectrum of component Bb in our data, which we 
ignored in the present analysis. To take this additional contribution into 
account and to properly analyze triple-lined 
systems, we are developing a three-dimensional correlation algorithm (Mazeh et 
al. in prep.). Apart from enabling us to extract all the information
contained in our composite spectra, this new algorithm should prove useful in 
identifying the main factor that limits our current precision on the 
velocities of HD\,188753\,A.

While three planets have been found so far in binaries with a 
separation of $\sim$$20$~AU (Sect.~\ref{introduction}), HD\,188753\,Ab was the 
only planet known to reside in a tighter system. If HD\,188753\,Ab is removed 
from the list of planetary candidates, it may be tempting to associate 
the value of $\sim$$20$~AU with a ``minimum separation'' for considering that 
a binary (or a pair) possibly harbors a circumprimary giant planet. 
Nevertheless, we must bear in mind that selection 
effects are still strongly against planet detection in the closest binaries. 
Hence, the ``limit'' at $\sim$$20$~AU may alternatively reflect our present 
detection capabilities, or rather their limits. 
In order to investigate the occurrence of planets in the closest binaries, we
must therefore first characterize our detection capabilities in detail for 
various types of systems. This work is in progress, and 
definitive results from the two planet search programs targeting spectroscopic 
binaries \citep{Eggenberger02,Konacki05b} should provide stronger constraints on
the reality of the 20-AU ``limit''.


\begin{acknowledgements}
      We thank X.~Delfosse and F.~Galland for scheduling arrangements and for 
      carrying out some of the observations of HD\,188753. 
      We warmly thank F.~Bouchy, R.~da~Silva, B.~Loeillet, C.~Moutou, F.~Pont, 
      N.~Santos, and D.~S\'egransan for carrying out observations of 
      HD\,188753 during their observing runs. 
      We thank S. Zucker and B. Markus for their involvement and help in the  
      development of the multi-order version of TODCOR. 
      We thank Y. Tsodikovich for performing the least-square fit in the
      Wilson-Mazeh plot. 
      We acknowledge support from the Swiss National Research Foundation 
      (FNRS), the Geneva University, and the Israeli Science
      Foundation through grant no. 233/03.  
      This work has made use of the ORBIT code developed by T. Forveille 
      \citep{Forveille99}. 
      
\end{acknowledgements}


\bibliographystyle{aa} 
\bibliography{6835bib}

\end{document}